\input psfig.tex
\documentstyle[]{mn}

\def\RX{RXJ1340.6+4018~}
\def\deg{\hbox{$^\circ$}}

\def\sol{\mbox{$_{\odot}$}} % solar-mass symbol

\def\erg{{\rm\thinspace erg}}
\def\cm{{\rm\thinspace cm}}
\def\km{{\rm\thinspace km}}
\def\s{{\rm\thinspace s}}
\def\ergps{\hbox{$\erg\s^{-1}\,$}}
\def\ergpcm2ps{\hbox{$\erg\cm^{-2}\s^{-1}\,$}}
\def\kmps{\hbox{$\km\s^{-1}\,$}}
\def\h50{\hbox{h$_{50}$}}
\def\m12{\hbox{$\Delta$m$_{12}$}}
\def\lesssim{\mathrel{\hbox{\rlap{\hbox{\lower4pt\hbox{$\sim$}}}\hbox{$<$}}}}
\def\gtrsim{\mathrel{\hbox{\rlap{\hbox{\lower4pt\hbox{$\sim$}}}\hbox{$>$}}}}

\title[An endpoint of galaxy merging]
{Multi-wavelength observations of an evolved galaxy group: 
An endpoint of galaxy merging? }
\author[Jones, L.R., Ponman, T.J. \& Forbes, D.A.]
{
L. R. Jones, T. J. Ponman and Duncan A. Forbes\\
School of Physics \& Astronomy, University of Birmingham, 
Birmingham B15 2TT, UK.\\
}
\begin{document}
\maketitle

\begin{abstract}
The group of galaxies \RX has approximately 
the same bolometric X-ray luminosity
as other bright galaxy groups and poor clusters such as the Virgo cluster. However,
70\% of the optical luminosity
of the group comes from a dominant giant elliptical galaxy, 
compared to 5\% from M87 in Virgo. 
%despite the
%similarity of the optical luminosities of the dominant elliptical and M87.
%Other X-ray properties of \RX are similar 
%to those of well known galaxy groups, but the distribution of optical
%galaxy luminosities is very different. 
The second brightest galaxy in \RX is a factor
of 10 fainter (\m12=2.5 mag) than the dominant elliptical, and the 
galaxy luminosity function has a gap at about L$^*$. 

We interpret the properties of the system as a result of galaxy merging
within a galaxy group. We find that the central galaxy lies on the 
fundamental plane of ellipticals, has an undisturbed, non-cD morphology, 
and has no spectral features indicative of recent star formation,
suggesting that the last major merger occurred $\gtrsim$4 Gyr ago.
The deviation of the system from the cluster $L_X-T$ relation in the
opposite sense to most groups may be due to an early epoch of
formation of the group or due to a strong cooling flow.

The unusual elongation of the X-ray isophotes and the similarity
between the X-ray and optical ellipticities at large radii ($\sim$230 
kpc) suggest that both the X-ray gas and the outermost stars of the dominant
galaxy are responding to an elongated dark matter distribution.
\RX may be part of a filamentary structure related to infall in
the outskirts of the cluster A1774.
\end{abstract}

\begin{keywords}
galaxies: clusters: general - X-rays: galaxies - galaxies: elliptical
\end{keywords}

\section{INTRODUCTION}

If galaxy merging is a common process in the cores of galaxy groups, 
% as predicted in hierarchical models of structure formation (), 
then there
may be a population of `fossil' groups in which most of the galaxies
have merged, leaving a giant elliptical galaxy but few other galaxies.
Galaxy groups represent one of the environments where galaxy mergers are
predicted to happen most frequently, because of the relatively low
velocity dispersion of the constituent galaxies compared to the 
velocity dispersion
in clusters of galaxies. For example, in the numerical
simulations of Barnes (1985, 1989) compact group members merge to form a
single elliptical galaxy within a few billion years. 
%If galaxy merging
%is an important phenomenon, then its influence should be detectable in
%isolated galaxy groups which, 
In  hierarchical models of structure
formation, isolated galaxy groups are predicted to be relatively
old systems, so that there has been a long period over which merging could occur.
%be an important process since it has

%In the hierarchical model of structure formation, 
%there should be a higher fraction of groups in the fossil stage than rich
%clusters, since infall of galaxies 
%into rich clusters is predicted to continue today.

The first candidate fossil galaxy group, RXJ1340.6+4018, was discovered
via its extended X-ray emission by Ponman et al (1994; hereafter P94).
The X-ray properties of the system, including its bolometric X-ray
luminosity (4.5x10$^{43}$ erg s$^{-1}$), temperature (0.92$\pm$0.08 keV)
and extent (core radius 140-390 kpc), are similar to those of well known
X-ray bright galaxy groups (e.g. Mulchaey et al 1996). The X-ray luminosity and extent
are at least an order of magnitude higher than those of normal elliptical
galaxies.  However, P94 detected no excess of galaxies in projection on
the sky near the X-ray position; only a single luminous ($M_V$=--23.5 mag)
elliptical galaxy was coincident with the X-ray peak. The 
explanation suggested by P94 was that the former group galaxies had
merged, creating the single 
luminous elliptical galaxy and leaving a halo of X-ray
emitting gas and dark matter.

In a recent study, Vikhlinin et al (1999) find three
very similar systems, in addition to independently discovering \RX, and
find that the space
density of such groups is similar to that of
luminous field ellipticals with $L>6L^*$ (M$_R<$--22.5).
Thus a large fraction of luminous field ellipticals
may be the result of galaxy merging in groups.
X-ray observations of an optically-selected isolated elliptical
galaxy by Mulchaey \& Zabludoff (1999) also revealed a system with the 
X-ray properties of a group, lending
support to this idea.

In this paper we investigate the properties of RXJ1340.6+4018 in more detail and
assess the evidence for it being a result of galaxy merging. 
If it {\it is} an old merger remnant 
%(and this may be one of the few ways of identifying
%such a system) 
then comparison with typical ellipticals gives a valuable check on
the idea that many ellipticals are the result of mergers.
The detection of an overluminous X-ray halo containing only one bright
(elliptical) galaxy may be one of the few ways of identifying such 
evolved systems. 

In Section 2 we describe new optical imaging and spectroscopy 
and high resolution X-ray observations of the \RX field. The results of 
isophotal, photometric and dynamical analyses of the central galaxy 
and its environment are described in Section 3. A comparison with
other galaxy systems and a discussion of the possible history of 
this system are given in Section 4. The last Section contains
the concluding remarks. We assume H$_0$=50\h50 \kmps Mpc$^{-1}$ and
q$_0$=0.5 throughout (as done by P94).

\section{Observations}

\subsection{Optical Observations~}
\subsubsection{Redshifts of nearby galaxies}
Low resolution spectra of RXJ1340.6+4018 and objects close to it on the sky
were obtained at the Kitt Peak 2.1m telescope on January 21st 1996.
A finding chart is shown in Fig. 1. Two slit positions were used.
At one, spectra of objects G2, G9, G7 \& S4 were obtained
(position angle 79\deg; exposure 60 min). The star S4 is outside the 
area of Fig. 1, to the east. At the other slit position
spectra of objects G1, G3 and G5 were obtained (position angle 71\deg;
exposure 30 min). 
Object G1 is the central, luminous elliptical
galaxy. The second slit was aligned on galaxies G3 \& G5
and so did not quite cross the core of G1, but passed 
2 arcsec south of the core.
The Goldcam spectrograph and Ford 3kx1k CCD were used with 
a slitwidth of 2.5 arcsec
and a grating of 300 lines/mm, giving a resolution of 
10\AA ~(FWHM). Arc lamp, flat field and F8 star exposures 
(the latter for the removal of
atmospheric absorption features) 
 were obtained with the telescope at or near the position of
RXJ1340.6+4018. Spectra of flux standard stars were obtained during the night. 
Standard IRAF reduction procedures were used and
redshifts were determined from Gaussian fits to
4-8 absorption lines per galaxy. Redshift errors were
determined from the dispersion in the individual line measurements. 
The redshifts are listed in Table 1. 

The redshift of G1 (z=0.1716) agrees with the measurement of P94.
Three of the other galaxies observed (G2, G3 and G5) have redshifts 
very close to that of G1 ($\Delta$v$<$450 km s$^{-1}$). The spectra of all 
four of these galaxies are typical of elliptical galaxies. 
Two objects (G7 \& G9) are background galaxies and one object (S4) is
a Galactic star. The only emission lines detected
were unresolved [OII]3727 in the background galaxies G7 and G9, a
common feature in faint field galaxies.

\begin{table}
%\begin{minipage}{175mm}
\caption {Redshifts of galaxies in the \RX field}
 \begin{tabular}{lllll} \hline
Galaxy & $\alpha$ (J2000)& $\delta$ & z & R $^{\dag}$ \\
 & & & &  \\
G1 & 13 40 32.80 & 40  17  39.7 & 0.1716 $\pm$ 0.0003 & 16.1 \\
G2 & 13 40 32.49 & 40  18   9.9 & 0.1705 $\pm$ 0.0002 & 18.6 \\
G3 & 13 40 31.27 & 40  17  31.2 & 0.1700 $\pm$ 0.0006 & 18.9 \\
G5 & 13 40 26.64 & 40  17  13.1 & 0.1734 $\pm$ 0.0006 & 19.2 \\
G7 & 13 40 39.25 & 40  18 25.4  & 0.3033 $\pm$ 0.0002 & 18.7 \\
G9 & 13 40 37.43 & 40 18 22.2 &   0.3087 $\pm$ 0.0004 & 19.2 \\
 & & & &  \\ 
             \hline
   \end{tabular}

$^{\dag}$ total magnitude\\
%\end{minipage}
\end{table}

\subsubsection{Internal velocity dispersion of the luminous galaxy}
A higher resolution, high signal-to-noise spectrum of the brightest
galaxy G1 was obtained at the Keck II telescope on May 2nd 1998.
The observation was made through thick cloud and so no flux
standards were observed. The low resolution imaging  spectrograph (LRIS), 
a grating of 
600 lines/mm and a slitwidth of 1.5 arcsec were used to give a resolution
of 6.7\AA ~(FWHM), corresponding to $\sigma\approx$140 \kmps.
 The slit was aligned across the galaxy core at 
a position angle of 90\deg. Arc lamp and flat field exposures were
obtained immediately after the target exposure. Standard IRAF reduction
procedures resulted in the spectrum shown in Fig. 2 which was
extracted from  a 2.2 arcsec section along the slit, centred on
the galaxy core. The signal-to-noise in the reduced spectrum is $\approx$30
per 1.2\AA ~pixel. 

The redshift of z=0.1724$\pm$0.0002, derived from Gaussian
fits to 9 absorption lines,
is consistent (within 2$\sigma$) with that derived from the 
Kitt Peak 2.1m spectrum.  The velocity dispersion in the galaxy core was
measured from Gaussian fits to the four absorption lines which had
equivalent widths $>$1\AA ~but were not blended with nearby strong
lines (a Fourier analysis was not possible because no template stars 
were observed with the 
same grating). These four lines were Fe 4383, Ca 4455, Fe 5270 and
H$\beta$. The observed weighted mean line width was $\sigma$=320$\pm$
40 \kmps.
The strong NaD lines could not be used because they
were contaminated by the O$_2$ B-band atmospheric absorption feature.
Although the Mgb triplet was not resolved, a fit with 
three fixed line centres and a single variable line
width and a variable continuum gave a line width ($\sigma$=350$\pm$30 \kmps)
consistent with the 
results from the other four lines. Errors on the line widths were 
estimated via 100 Monte-Carlo simulations for each line. 
After correcting for the instrumental broadening, the weighted mean
line width was $\sigma_0$=260$\pm$30 \kmps.

\begin{figure}
\vspace{10cm}
\caption{R band CCD image of the central region of the
group \RX, showing the giant
elliptical galaxy G1 and other galaxies in the field. Galaxies G2, G3
\& G5 are in the galaxy group dominated by galaxy G1; G7 \& G9 are
background galaxies. The faint galaxy G10, 9\arcsec ~W of the centre of G1, 
is unusually blue (its redshift is
unknown). 
A bright star is marked with an "S". North is up and
east is left. The image is 2.5 arcmin across (570 \h50 kpc at the
redshift of \RX). The intensity scaling is log(square root).
}
\end{figure}

\begin{figure}
\psfig{figure=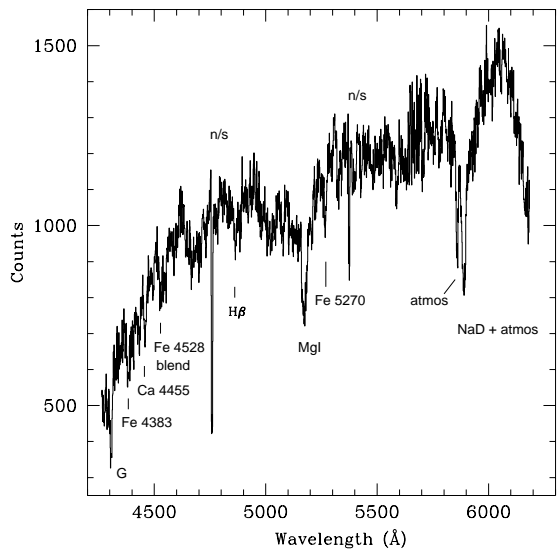,height=8.0truecm,angle=0}
\caption{Spectrum of the dominant galaxy G1 shifted to rest
wavelengths. Prominent absorption lines and incompletely subtracted night sky 
lines are marked.
}
\end{figure}

\subsubsection{Deep optical imaging}
Images of RXJ1340.6+4018 in the R and B bands were obtained in service
time at the Isaac Newton 2.5m Telescope (INT) on January 25-26th 1995. A Tektronix
1024x1024 CCD was used at prime focus, resulting in a pixel size of
0.59 arcsec and field of view of 9.8x9.8 arcmin. The seeing 
was 1.2 arcsec FWHM. Exposures of 6x600s in R
and 3x500s in B were obtained over the two nights, moving the telescope slightly between 
exposures to facilitate flat-fielding. A flat-field 
image was initially made from a combination of 6 twilight sky images.
Application of this flat-field image to the target images left a
residual gradient in the sky of $<$0.2\% over the full 9.8 arcmin frame width. 
The three frames with the
worst residual gradient (which were those for which no twilight flats
were obtained on the same night as the target data) were additionally
flat-fielded by a smoothed `sky flat' created by a median combination
of the three original frames. This operation was 
designed to remove
the large scale gradient. The sky in the resulting images was flat to within $<$0.1\% 
across the full extent of each  frame (corresponding to R$>$28 
mag arcsec$^{-2}$).
The final R band image from the six frames combined is shown in Fig. 3.

The INT images were obtained in slightly non-photometric conditions, so
photometric calibration 
was performed via the Nordic Optical 2.5m Telescope (NOT) images of the same field
described in P94
in the R band, but was unavailable for the B band. Here we give a 
brief description of the V and R photometry employed at the NOT. The 
photometry was on the Johnson (V) and Kron-Cousins (R) system.
The NOT images were
not as deep as those described here, but were obtained under 
photometric conditions, as measured by 
12 observations of Landolt (1992) and Christian et al (1985) standard star
fields in V and R throughout the night. These standard star 
measurements were consistent with standard atmospheric extinction corrections
of $0.06sec(z)$ in R and $0.10sec(z)$ in V, where z is the airmass.
Colour terms of $0.0(V-R)$ in R and $0.34(V-R)$ in V were derived from
17 stars (in R) and 4 stars (in V). Several standard stars were of
very similar V-R colour to galaxy G1. We estimate the NOT photometry
to be accurate to 0.05 mag. Calibration of the deep INT image
was achieved via a simple comparison with the NOT R photometry for
the brightest 8 galaxies (including those with measured redshifts)
within a 6.6 arcsec radius aperture, and was accurate to 0.03 mag 
(1 $\sigma$ random error).

\begin{figure}
\vspace{10cm}
\caption{R band CCD image and overlaid X-ray contours from the ROSAT PSPC and 
ROAST HRI (dashed).
The X-ray contours are at logarithmic 
intervals of a factor of 1.4 in surface brightness. 
}
\end{figure}

\begin{figure*}
\begin{minipage}{125mm}
\vspace{8cm}
\caption{Azimuthally averaged X-ray surface brightness profiles. 
X-ray emission is 
detected to a radius of $\gtrsim$ 0.1$\deg$ or 1.3 \h50$^{-1}$ Mpc.
$ROSAT$ PSPC 
data are shown as crosses and scaled HRI data as points. The PSPC/HRI count rate ratio
used to scale the HRI data was predicted using the spectral parameters measured from
the PSPC spectrum.
}
\end{minipage}
\end{figure*}

\begin{figure}
\vspace{10cm}
\caption{The same field as shown in Fig. 3, after subtraction of an elliptical model fit
to galaxy G1 as described in the text. There are few, if any, residual
features.
}
\end{figure}

\subsection{X-ray Observations}
RXJ1340.6+4018 was observed with the ROSAT High Resolution Imager (HRI) 
for a total of 51 ksec in December 1994 and June 1995. 
The HRI count rate was 2.9x10$^{-3}$ ct s$^{-1}$ within a radius of 1.2
arcmin. 
The HRI has a spatial resolution of 5 arcsec (FWHM), and
is reasonably sensitive to point-like X-ray sources, but 
because of a higher unrejected particle background and lower quantum
efficiency, it is not as 
sensitive to faint diffuse X-ray emission as the ROSAT position sensitive
proportional counter (PSPC) which was used by P94. Nevertheless,
an extended source was detected in the HRI at the position of the 
peak of the PSPC emission (see the dashed contours in Fig. 3). The 
position of the HRI image 
was updated by 5 arcsec by comparing the X-ray positions of 5 point
sources in the HRI image with the positions of optical counterparts from
Palomar plate measurements.
The PSPC data is subject to a systematic
positional uncertainty 
of $\approx$10 arcsec.
No point X-ray sources were detected in the
HRI image within the PSPC contours of Fig. 3, showing that the 
extended emission detected with the PSPC by P94 (with a poorer 
spatial response FWHM$\approx$25 arcsec) is from a truly
diffuse source, and not from a superposition of unresolved point 
sources. Any non-variable point source with a 
flux of 10\% of the total flux of \RX would have been detected 
with a significance of at least 5$\sigma$ in the HRI observation.

The combined HRI and PSPC azimuthally averaged 
surface brightness profiles are shown in Fig. 4. X-ray emission is 
detected to a radius of $\gtrsim$ 0.1$\deg$ or 1.3 \h50$^{-1}$ Mpc.
In the figure, the HRI surface brightness has been scaled up by a factor
of 3.78, corresponding to the predicted PSPC/HRI count rate ratio for a 
thermal plasma spectrum as determined by P94 (T=0.92 keV, metallicity=0.36
solar, N$_H$=7.4x10$^{19}$ cm$^{-2}$). The PSPC surface brightness profile
is consistent with the HRI profile in the overlap region, and consistent
with a smooth extrapolation of the HRI profile, again suggesting that 
unresolved point sources do not make a significant contribution to the
PSPC flux.

\section{Analysis and Results}

\subsection{The nature of the system}

The luminous elliptical galaxy coincident with the peak of the 
X-ray emission is not completely isolated, but has at least three faint 
companion galaxies with
measured redshifts. The system is a group of galaxies, and the diffuse 
X-ray emission originates in
an intra-group medium (IGM). The group, however, has an unusually
optically luminous central galaxy and an unusually large magnitude
gap between the brightest and second brightest galaxies. 
Here we determine the properties of both 
the central galaxy and the galaxy group. The group properties are
summarized in Table 2.

\begin{table*}
\begin{minipage}{125mm}
\caption {Properties of the \RX group}
 \begin{tabular}{ll} \hline

Redshift & z=0.1710$\pm$0.0007 \\
Velocity dispersion & $\sigma_r$=380 (+350, -110) \kmps \\
Bolometric X-ray luminosity$^{\dag}$ & 4.5x10$^{43}$ \h50$^{-2}$ \ergps \\
X-ray temperature$^{\dag}$ & T=0.92$\pm$0.08 keV \\
IGM X-ray metallicity$^{\dag}$ & Z=0.36 (+0.15,-0.11) solar \\
Magnitude gap (in R)& \m12=2.5 mag  \\
Central galaxy surface density & N$_{0.5}^c$= 2.8\\
Total gravitating mass (r$<$340 kpc)$^{\dag}$ & 2.8x10$^{13}$ \h50$^{-1}$ M$\sol$ \\
Virial mass$^{\ddag}$ & 6.1x10$^{13}$ \h50$^{-1}$ M$\sol$ \\
\hline
 \end{tabular}

$^{\dag}$ from Ponman et al (1994)\\
$^{\ddag}$  The virial mass is poorly constrained because of the small
number of measured galaxy velocities.
 
\end{minipage}
 \end{table*}

\subsection{Group luminosity function}

The luminosity function of the galaxies in the group was estimated
using a simple statistical method. The surface density of field galaxies 
surrounding the group was subtracted from 
the surface density of galaxies located in a circular region centred
on the group, as a function of magnitude.
Photometry of both field and group galaxies was
based on analysis of the deep R band INT CCD image using the Starlink PISA package
(Draper \& Eaton 1996). The image used for source detection was that with the bright
elliptical galaxy subtracted (Fig. 5; see Section 3.4), so as to avoid spurious faint
sources being produced in the noisy, low surface brightness outer parts of the 
galaxy profile, and to include sources projected on to the bright
galaxy. Regions around four bright stars were also excluded from the
analysis, again to avoid spurious source detections. An isophotal 
threshold of 26.6 R mag arcsec$^{-2}$ (2$\sigma$ per pixel above sky) was used, 
with a minimum of 4 connected pixels above the threshold, to define
source detections. A total of 2577 objects were found in an area
of 55.1 arcmin$^{2}$. The bright central galaxy was inserted manually into the
object list.

Photometry was performed using an asymptotic curve of
growth analysis for each object as described by Kron (1980). Star-galaxy 
separation was performed,
following Jones et al (1991),
using a log(area)-magnitude plot for R$<$18 and a
log(I$_{peak}$)-magnitude plot for 18$<$R$<$26, where the area is
the isophotal area above the threshold, and I$_{peak}$ is the peak 
surface brightness. The completeness level was estimated from 
galaxy counts in a background region, defined as 
the area of the image $>$1.8 arcmin (400 \h50$^{-1}$ kpc) 
from the centre of the galaxy G1. The background galaxy counts start to turn over
at R=24 mag, and we take this as the completeness limit (see Fig. 6).
The background galaxy counts are consistent with 
field galaxy counts of Jones et al (1991) to
their limit of R=22, and slightly higher than the  measurements of
other authors at 22$<$R$<$24 
(but within 2.1 times the one-sigma fluctuation amplitude measured 
by Metcalfe et al 1991 in a similar size field). 
This small excess may be due to the effects of the large-scale structure in the
region (perhaps associated with Abell 1774).

Our background galaxy counts predict 207 field galaxies 
within a 1.8 arcmin radius circle at R$<$24; we measure 250
galaxies within the same radius circle 
centred on galaxy G1, giving an excess of 43$\pm$16 galaxies
within a radius of 400 \h50$^{-1}$ kpc.  Within  200 \h50$^{-1}$ kpc radius
the excess is 26$\pm$8.8 galaxies. Using the field galaxy counts of
Metcalfe et al (1991, 1995) would increase the number of group
galaxies within 400 \h50$^{-1}$ kpc to 71$\pm$16, an increase
of a factor of 1.7, and an indicator of the likely level of systematic
error in the luminosity function.

K-corrections for an elliptical galaxy
at z=0.171 of K$_B$=0.84, K$_V$=0.32 (Pence 1976) and 
K$_R$=0.17 (Metcalfe et al (1991); Saglia et al (1997)) were assumed.
A Galactic extinction of A$_B<$0.06 mag was obtained from Burstein
\& Heiles (1984). Converting the neutral hydrogen column density of
N$_H$=7.4x10$^{19}$ cm$^{-2}$ to an extinction using the relation
of Jenkins \& Savage (1974) gives A$_B$=0.04 mag (or  A$_R$=0.02 mag).
As these values are less than our photometric errors, 
we have assumed that the Galactic extinction is zero.

The group luminosity function is shown in Fig. 7. 
The luminosity function was constructed 
using the background counts derived from the same image as the
group counts, in order to have consistent depth, galaxy detection 
parameters and photometry. A bin size of one magnitude has been
used in Fig. 7.
The large magnitude gap \m12 between the 
brightest galaxy and the second brightest galaxy is clear in this
figure (the value is \m12 = 2.5 mag). The unusually high
luminosity of the brightest galaxy when compared to the  LFs of
other clusters and groups is also evident. 

The LFs of other systems shown in Fig. 7 are representative of the
LF {\it shape} in groups and poor clusters. Their detailed normalization
should not be compared because they have been measured from different
projected areas in each system. In addition, the Virgo LF has been 
rescaled by a factor of 0.1 in order to compare its shape with the LF
of \RX.

\begin{figure}
\psfig{figure=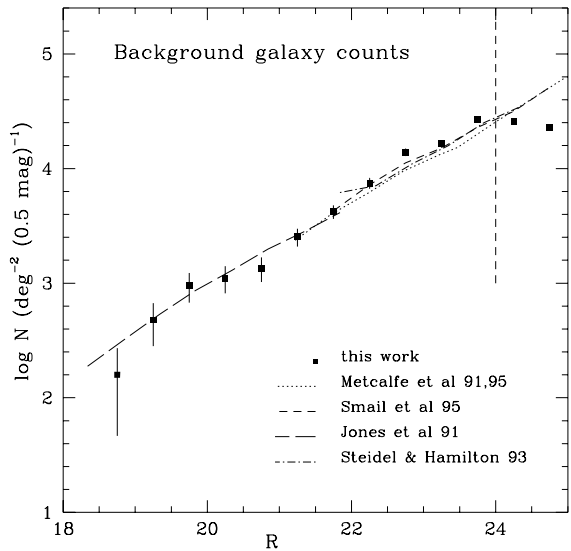,height=8.0truecm,angle=0}
\caption{Background galaxy counts at radii $>$400 \h50$^{-1}$ kpc from
the centre of the \RX group. The limit of completeness is shown by the vertical
line. Field galaxy counts of other authors are shown as dashed lines.
The counts of Steidel \& Hamilton (1993) have been transformed to the R band
using R=$\cal R$-0.14.
}
\end{figure}

\begin{figure}
\psfig{figure=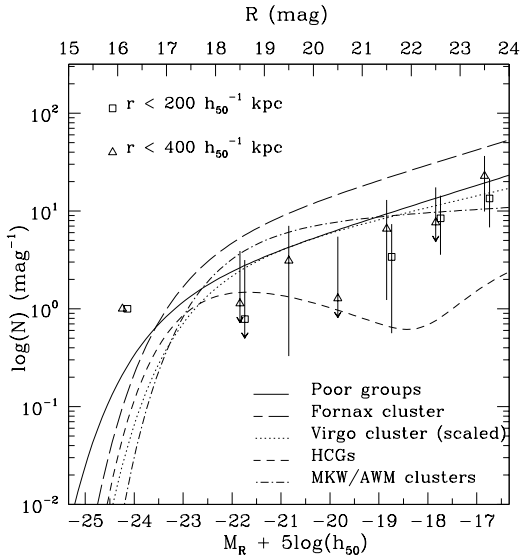,height=8.0truecm,angle=0}
\caption{Galaxy luminosity function of the \RX group, for galaxies
within two different radii (shown as triangular and square symbols). 
The large magnitude gap between the brightest galaxy
(G1;
M$_R$=-24.2) and the second brightest (M$_R$=-21.7) is evident, as is the low likelihood
of other clusters \& groups containing a galaxy as luminous as this. 
The other luminosity 
functions (shown as continuous lines) 
are taken from Ferguson \& Sandage (1991) except for
the Hickson compact group (HCGs) LF which is taken from Hunsberger, Charlton \&
Zaritsky (1998) (scaled to H$_0$=50) and the MKW/AWM poor cluster LF
which
is taken from Yamagata \& Maehara (1986). Colour transformations to
the R band assumed  B-R=1.5 mag 
for the Ferguson \&
Sandage results and V-R=0.6 for Yamagata \& Maehara.
Only the Virgo cluster LF has been scaled in number (by a factor of 0.1).
}
\end{figure}

\subsection{Group velocity dispersion and dynamical mass}

The redshift and line-of-sight velocity dispersion ($\sigma_r$) 
of the group of galaxies was estimated from
the four available galaxy redshifts. For small sample sizes, Beers et al. (1990) found
that the median redshift and simple Gaussian dispersion give the best estimates.
Adopting this method gives a group 
redshift of z=0.1710$\pm$0.0007 and a velocity dispersion 
$\sigma_r$=380 (+350,-110) km s$^{-1}$
where the velocity dispersion
has been corrected for cosmological effects and redshift measurement errors, and its
error has been calculated as described by Danese et
al. (1980). Zabludoff \& Mulchaey (1998) note that group velocity dispersions
measured from a small number of centrally concentrated  brightest galaxies usually
underestimate the true velocity dispersion, as measured from a much
larger sample, by a factor $\sim$1.5. 
The kinematic quantities derived from the velocity dispersion are of dubious value, because
they were based on only 4 galaxy velocities and positions, and those 4 galaxies were 
concentrated towards the centre of the group.
With this caveat,
we calculate a virial mass using the standard equations as given by Ramella et
al (1989). 
For a virial radius of 310 kpc and a
mean projected galaxy separation of 330 kpc, we find a virial mass of 
6.1x10$^{13}$ M$\sol$.

\subsection{Isophotal analysis and photometry of the central galaxy}

In order to determine if the central elliptical galaxy G1 shows any
deviations from an elliptical shape and to measure its photometric
properties, we used the {\small ISOPHOTE} package in {\small IRAF} to fit 
elliptical isophotes to the deep R band image, using the algorithm of
Jedrzejewski (1987). The regions around several stars and galaxies 
close to G1 on the sky were ignored in the fitting. The galaxy profile
could be followed to a maximum semi-major axis length (a) of 62 arcsec or
230 \h50$^{-1}$ kpc. The surface brightness values were corrected for 
K-corrections and (1+z)$^4$ surface brightness 
dimming. 
The R band image after subtraction of the best-fit elliptical model
is shown in Fig. 5. This image reaches a limiting surface brightness of 
$\mu_R>$26 mag arcsec$^{-2}$, considerably fainter than,
for example, the tidal tails in NGC 7252, a 1 Gyr old merger remnant 
($\mu_R\approx$25 mag arcsec$^{-2}$;
Hibbard et al 1994). However, no diffuse residual features are visible in this
image; only a few galaxies and a stellar object near the galaxy centre
are visible. 
While the two features closest to the galaxy centre may be 
multiple nuclei, they may also be a faint galaxy and a star 
observed in projection against the central galaxy. The high number density
of similarly faint objects throughout the image suggests that they 
may be unrelated to the central galaxy.
The small feature at
the galaxy core is an artifact of the imperfect fit in the central
region due to seeing effects. 

The surface brightness as a function of a$^{1/4}$
is shown in Fig. 8. The straight line, indicative of a de Vaucouleurs
r$^{1/4}$ profile with an effective semi-major axis of
a$_e$=10.6$\pm$0.6 arcsec, 
is a good fit in the range 3.0$<$a$<$54 arcsec, ie. outside the central 
region affected by the seeing ($\chi^2$=18.0 for 23 degrees of freedom). 
An effective radius, r$_e$, of
6.4$\pm$0.3 arcsec, is obtained from a de Vaucouleurs profile fit
to r=$\sqrt{ab}$ over the range 3.0$<$r$<$36 arcsec. This value of
r$_e$ is consistent with that found by P94. The B band image is of
lower signal-to-noise than the R band image, but following the
same procedure again gave a profile consistent with a r$^{1/4}$ law,
with a$_e$=13.8$\pm$3.0 arcsec and r$_e$=7.8$\pm$0.9 arcsec.

The variation of ellipticity with semi-major axis is given in
Fig. 9. Here ellipticity is defined as e=1-b/a. The ellipticity
increases monotonically with radius from e=0.1 (ie. near circular) to e=0.5
(axis ratio b/a of 0.5). This range of ellipticity within individual 
luminous elliptical galaxies is not unusual, although galaxies with 
similarly large ranges of ellipticity also usually have disk or dust lane
features which affect the isophotal analysis (eg. Goudfrooij et al 1994). The position
angle of the major axis is approximately constant between 20\deg and 25\deg. 
The amplitude of the 4th cosine harmonic measures `boxy' or `disky'
deviations from perfect ellipticity in the isophotes. 
The value of this amplitude is 
consistent with zero at all radii outside the central seeing-affected
region, implying that no significant deviations are present, at least in our data.

Total magnitudes of the galaxy were measured by summing the light
within a radius of 5.6r$_e$ and adding a correction of 0.1 mag given
by the r$^{1/4}$ profile for the light outside this radius. 
The total magnitudes are given in Table 3.

No photometric calibration was available for the B band. An 
approximate calibration was performed by assuming that the rest
(B-V)$_0$ colour of galaxy G1 (at r$<$16 arcsec) was (B-V)$_0$=0.95,
given by the colour-central velocity
dispersion relation for elliptical galaxies of Forbes, Ponman \& Brown
(1998).

\begin{table*}
\begin{minipage}{125mm}
\caption {Properties of the central elliptical galaxy G1}
 \begin{tabular}{ll} \hline

 total magnitude & R=16.1  \\ 
                 & M$_R$=-24.2 + 5log(\h50)\\
                 & M$_V$=-23.4 + 5log(\h50) \\
 type            & E5 \\
 rest-frame colour (r$<$50 kpc) & (V-R)$_0$=0.72$\pm$0.06  \\
 central velocity dispersion & $\sigma_0$=260$\pm$30 \kmps \\
 effective radius (R) & r$_e$=6.4$\pm$0.3 arcsec = 24.1$\pm$1.1
\h50$^{-1}$ kpc \\
 effective radius (B) & r$_e$=7.8$\pm$0.9 arcsec = 29.4$\pm$3.4
 \h50$^{-1}$ kpc \\
 effective semi-major axis (R) & a$_e$=10.6$\pm$0.6 arcsec =
40.0$\pm$2.3 \h50$^{-1}$ kpc\\
 effective semi-major axis (B) & a$_e$=13.8$\pm$3.0 arcsec =
52.0$\pm$11.3 \h50$^{-1}$ kpc\\
 Mean surface brightness &   $\mu_R$=21.68 mag arcsec$^{-2}$ \\
 ~within a$_e$ (corrected) & $\mu_{r_G}$=21.37 $\mu_B$=23.65 
  mag arcsec$^{-2}$ $^{\ddag}$ \\
 Mg$_2$ index$^{\S}$  & 0.30$\pm$0.15 \\
 Radio power at 1.5 GHz$^{\dag}$ & 3.5x10$^{30}$ \ergps Hz$^{-1}$ \\

 \hline
 \end{tabular}

$^{\dag}$ from Ponman et al (1994)\\
 $^{\ddag}$ accurate to $\approx$0.3 mag in B.\\
 $^{\S}$ Mg$_2$ index as defined in Dressler (1984) and Faber et al
(1977).
\end{minipage}
\end{table*}

\begin{figure}
\psfig{figure=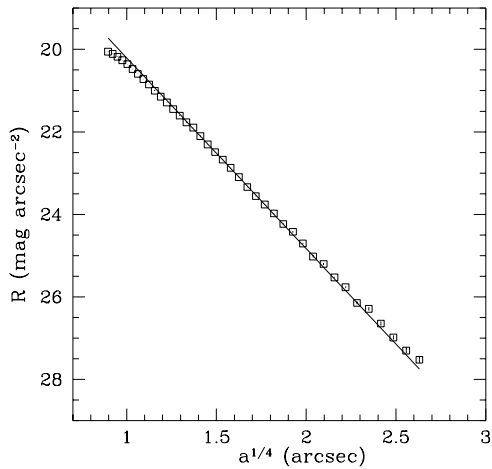,height=8.0truecm,angle=0}
\caption{R band surface brightness of the dominant galaxy G1 as a function of (semi-major
axis)$^{1/4}$, from a fit to elliptical isophotes. A pure de Vaucouleurs
r$^{1/4}$ profile (solid line) is a good description outside the central region
affected by the seeing.
}
\end{figure}

\begin{figure}
\psfig{figure=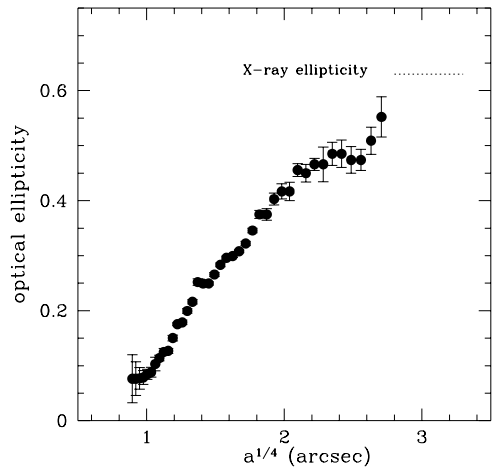,height=8.0truecm,angle=0}
\caption{Ellipticity of the dominant galaxy G1 as a function of
(semi-major axis)$^{1/4}$, from a fit to elliptical isophotes. The
ellipticity increases from near-circular at small radii to approach
the X-ray ellipticity, shown by a dotted line, at large radii.
}
\end{figure}

\subsection{Fundamental Plane and Scaling Relations}

If the luminous central galaxy is a result of recent multiple mergers, then
its photometric and dynamical properties may not have yet relaxed to
those of typical ellipticals. For normal ellipticals, these properties
are defined by the fundamental plane. 

In order to test if the central galaxy G1 lies on the fundamental plane of elliptical
galaxies, the central velocity dispersion, the effective
radius and the mean surface brightness are required. To compare 
with the
fundamental plane of Djorgovski \& Davis (1987), we used their definitions:
the effective semi-major axis a$_e$ from r$^{1/4}$ profile fits
and the mean surface brightness $<\mu>$ within a$_e$ in the Lick r$_G$ 
band. We used H$_0$=100 \kmps Mpc$^{-1}$ to be consistent with Djorgovski \& Davis (1987)
(as opposed to  the value of H$_0$=50 \kmps Mpc$^{-1}$
used throughout the rest of this paper).
The R band measurements (on the Kron-Cousins system) were converted
to the Lick r$_G$ band via the equations given in  Djorgovski (1985) and
Bessell (1979). The values found
(a$_e$=20$\pm$1.1 h$_{100}^{-1}$ kpc, $<\mu>_{r_{G}}$=21.37 mag arcsec$^{-2}$ and 
$\sigma_0$=260$\pm$30 \kmps) lie almost exactly on the fundamental plane 
relation of 
Djorgovski \& Davis (1987) and are certainly consistent with the galaxy G1
being on the fundamental plane within our measurement errors.

We tested for consistency using the B band fundamental plane parameters,
as given in Table 3. The galaxy was again found to lie on the fundamental plane,
defined in the B band by the data of Prugniel \& Simien (1996).

The Mg$_2$ absorption line index is an age/metallicity indicator. The index,
as defined by Dressler (1984) and Faber et al. (1977), is the 
depression, in magnitudes, of the intensity at wavelengths 
5157-5198\AA ~compared to continuum sidebands. The mean value from the KPNO 2.1m spectrum
and the Keck spectrum was 0.30$\pm$0.15.
This value is consistent with the elliptical Mg$_2$-$\sigma_0$ relation of 
Bender, Burstein \& Faber (1993), where $\sigma_0$ is the central velocity
dispersion. We also note that the measured rest-frame (V-R)$_0$ 
colour of the galaxy 
is consistent with the colour-magnitude relation of Prugniel \& Simien
(1996).

\subsection{Neighbouring galaxy properties}

If galaxy merging is an important process, then there may be signs
of it in galaxies close to the group centre.
The colours and mean surface brightnesses of galaxies at a range of projected
radii from the dominant galaxy were measured and are plotted in
Fig. 10. Only relatively bright galaxies (R$<$21.5) were investigated,
since colours were available for all of these galaxies in the area studied.
The colours were measured on images with the dominant elliptical
galaxy G1
subtracted, and using fixed apertures of 
4.7 arcsec diameter. The B-R colours are accurate relative to each
other, but are not accurate in an absolute sense because of the lack of 
B band calibration. 

Of the five galaxies within 200 \h50$^{-1}$ kpc of the 
centre of the dominant elliptical galaxy (shown as filled
squares in Fig. 10), three are outliers from the general distribution.
Galaxy G10 (identified in Fig. 1) is the bluest of all the galaxies
and has a low mean R band surface brightness. Its unusual appearance
can be best seen in Fig. 5. Ignoring galaxy G4,
which appears to be projected behind a foreground star, galaxy G3 has
one of the highest surface brightnesses. Both of these galaxies 
are very close (in projection) to the central galaxy; both are 
within its isophotes. In addition, galaxy G11 is very red.

Although we have no redshift for galaxy G10, its location only
9 arcsec (or 35 kpc) from the centre of the dominant galaxy, and its exceptional
colour and appearance argue strongly for an association between the
two galaxies. Its blue colour relative to other galaxies in the field
suggests recent star formation. This is not, however, a major merger.
Galaxy G10 emits $<$1\% of the R band light of the dominant galaxy.
One interpretation of the high surface brightness and small size of
the other unusual nearby galaxy, G3,
is that it is the high-density nuclear remains of a brighter galaxy.
In any case, the high fraction of abnormal galaxies close to the 
central galaxy suggests that interactions are occurring.

\begin{figure}
\psfig{figure=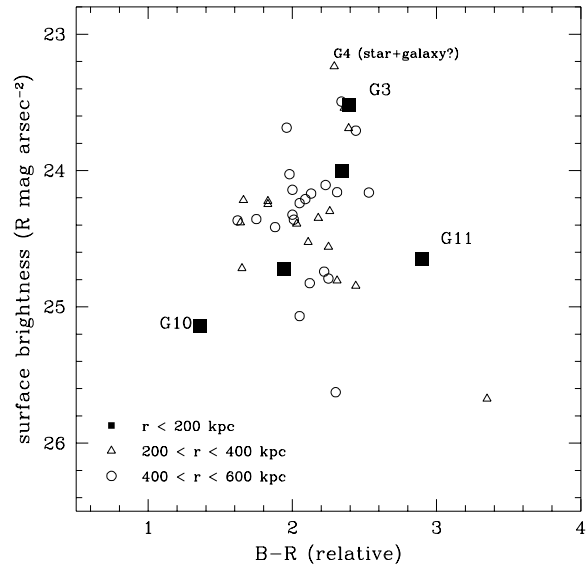,height=8.0truecm,angle=0}
\caption{The properties of galaxies close to the central galaxy. 
Mean surface brightness is plotted against relative B-R colour for 
galaxies with R$<$21.5 at a range of projected distances from the  central galaxy.
Three of the five galaxies within 200 kpc have fairly extreme
properties, suggesting galaxy interactions are occurring.
}
\end{figure}

\subsection{Central cooling flow}

The ROSAT HRI data allow an investigation of the 
properties of the hot X-ray emitting gas at the centre of the group,
coincident with the dominant galaxy.
We have performed a joint fit of all the X-ray data (PSPC spectral images
and HRI image) to a model
of the X-ray emitting gas, assuming hydrostatic equilibrium and
a King profile for the gas density
variation with radius  but with the
% temperature and 
metallicity fixed 
at the integrated value determined by P94. The spatial and spectral 
instrument response
functions were included as described in Ponman \& Bertram (1993).
The assumption of spherical symmetry is reasonable in the core of the 
group. We find that the cooling time of the X-ray gas decreases with
decreasing radius, and is below 10 Gyr for r$\lesssim$0.8 arcmin (190 \h50$^{-1}$
kpc). The cooling time is $\sim$1 Gyr for r$<$6 arcsec.
Using the method of White, Jones \& Forman (1997) we derive a mass 
deposition rate within r=0.8 arcmin
of 45 M$_{\sol}$ yr$^{-1}$ for a steady state cooling flow. However, 
we find no evidence of a strong 
decrease in the X-ray temperature at the group centre, as measured by
the PSPC. This could be due to the cooling flow being young, with
insufficient time for widespread cooling to set in after disruption by galaxy merging
perhaps 3-4 Gyr ago. However, the rather poor
PSPC spatial resolution ($\approx$25 arcsec) may be 
blurring the observed temperature profile such that, within the limited
statistical accuracy, the central temperature decrement is not observed.

\section{Discussion}

\subsection{Comparison with other groups}

\RX was selected from an X-ray survey.
Poor clusters selected optically on the basis of the D or cD-like
appearance of the brightest galaxy include those of Morgan, Kayser
\& White (1975, hereafter MKW) and Albert, White \& Morgan (1978,
hereafter AWM). The X-ray luminosity, X-ray temperature, 
velocity dispersion, mass estimates and mass/light ratio
of \RX all lie within the
range of the same parameters of the MKW/AWM poor clusters
(Kriss, Cioffi and Canizares 1983; Beers et al 1995; Bahcall 1980).
The distribution of the galaxy optical luminosities is, however, very
different in \RX. 

The extreme nature of the value of \m12=2.5 mag for \RX can be 
appreciated from the comparison with the distribution of \m12 in
poor clusters shown in Fig. 11. 
The comparison samples are the 25 optically selected poor clusters 
of Price et al (1991) and the 15 MKW/AWM clusters of Beers et al (1995)
with $\geq6$ galaxy redshifts. The mean of these distributions is 
\m12=0.4-0.5. Using a different sample of 16 MKW/AWM clusters,
and based on photometry alone, Bode et al (1994) quote a median
\m12 of 0.8 and a range of 0.1 to 1.8. The MKW/AWM clusters were
selected to contain a dominant giant elliptical galaxy, and thus
should have larger values of \m12 than randomly selected groups,
but even compared to this sample, \RX has an extreme \m12 value.

To quantify the random probability of obtaining a value of \m12$\geq$2.5,
we performed 14000
Monte Carlo realisations of luminosity functions with the 
absolute magnitudes
selected at random from a Schechter function distribution.
We used two of the LFs shown in Fig. 11 with extreme values of the
faint end slope, and 
a faint limit of M$_R$=-16, as in our observations.
Using the MKW/AWM composite poor cluster LF of Yamagata \& Maehara
(1986), which has a flat faint end slope ($\alpha$=-1.07) 
we found  that less than 0.1\% of the simulated
LFs had \m12$\geq2.5$. The mean and median \m12 values were 0.48 and 
0.38, and the 
\m12 distribution is shown in Fig. 11. For the composite poor
group LF of Ferguson \& Sandage (1991), with the steepest faint
end slope ($\alpha$=-1.39),
a larger fraction of galaxies had faint luminosities, resulting in higher
values of \m12 (1.2\% of the simulated
LFs had \m12$\geq2.5$). The mean and median \m12 values were 
0.71 and 0.55, less consistent with the observed distributions in
Fig. 11. 

To compare the simulated random probabilities with the number
of observed systems we use the results of Jones et al (1999) 
who find $\approx$5\% of 100 X-ray selected groups
and clusters in the WARPS survey have \m12$\geq2.5$, together
with the results of Vikhlinin et al (1999), who estimate from 4
similar systems that they comprise $\approx$20\% of all clusters 
and groups of comparable X-ray luminosity. These estimates are
significantly larger than the 0.1\%-1\% predicted from random
selection, suggesting that some other formation process is
at work.

If this other process is galaxy merging, how many L$^*$ galaxies
are required to form the dominant galaxy? Given the mean
M$^*_R$ of -23.0 for the LFs in Fig. 11, the answer is $\approx$3.
Three galaxies is also  consistent
with the number missing from the gap in the observed LF between the 
first and second brightest galaxies.

\begin{figure}
\psfig{figure=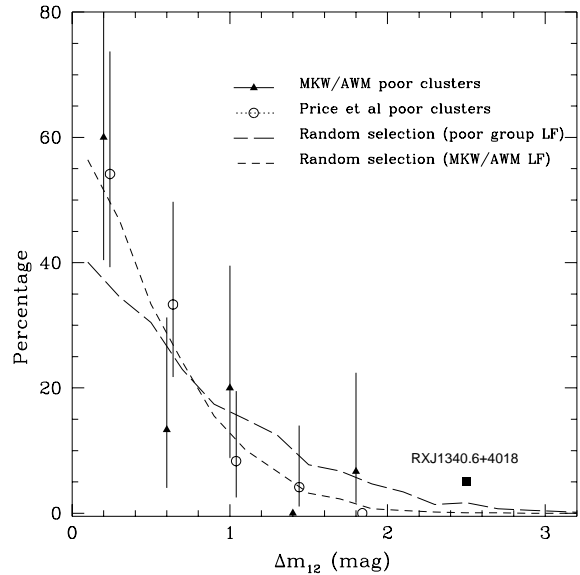,height=8.0truecm,width=8.0truecm,angle=0}
\caption{The distribution of \m12, the magnitude difference between the 
brightest and second brightest cluster  galaxies. The extreme value for
\RX can be compared with values calculated from the data of Beers et
al (1995) for 15 MKW/AWM poor clusters, and for the compilation of poor
clusters of Price et al (1991). The ordinate is the percentage per
bin of size 0.4 mag. Also shown are predictions based on random
selection from two of the Schechter function LFs shown in Fig. 7.
}
\end{figure}

A few of Hickson's compact groups (HCGs) have apparently large \m12 values
(eg. Fig. 11 of Prandoni et al 1994) but these are generally due to the projection
of unrelated foreground galaxies on to the groups; indeed HCGs 
were deliberately chosen to contain at least four galaxies of similar
magnitude (within 3 mag of the brightest galaxy), so the selection
criteria should bias them toward low values of \m12.

Related measurements are the fraction of the total group light in 
the  dominant
galaxy and the central galaxy surface density N$_0$. From the 
LF of Fig. 7 at r$<$400 \h50 kpc we find that $\approx$70\% of
the group light arises from the central galaxy. We have used a 
limiting absolute magnitude of M$_R$=--16 to calculate this fraction but
the result is valid for any absolute magnitude limit
fainter than M$_R\approx$--19 because the luminosity function integral 
converges.
By comparison, the fraction of the Virgo cluster light arising in M87
is only $\approx$5\%, as derived from the data of Sandage, Binggeli \& Tammann
(1985) and using a corresponding absolute magnitude limit. 
The Virgo cluster bolometric X-ray luminosity is
similar to that of \RX
(White, Jones \& Forman 1997).
The fractions of the cluster light emitted by the brightest cluster
galaxies in the MKW/AWM poor clusters have intermediate values bewteen
11\% and 40\% (from the data of Thuan \& Romanishin 1981), and a similar
range of values is found in the rich cluster data of Oemler (1976),
including clusters containing cD galaxies. Thus the fraction of light in
the dominant galaxy in \RX has an extreme value compared to a wide range of 
other systems.

The value of the central galaxy surface density, N$_{0.5}^c$, as defined 
by Bahcall (1981), was measured
as 2.8, assuming a richness correction appropriate for a richness
class -3 group. The value for the  Virgo cluster is 10 (Bahcall 1981).
Abell richness class 0-1 clusters typically have N$_{0.5}^c$=10-50.

We now make a comparison of \RX with 
Hickson compact groups.
\RX is more luminous in X-rays than any of the 85 Hickson compact groups
studied by Ponman et al (1996), and is a factor $\sim$15 more luminous
than the typical HCG detected. 
The group optical luminosity of \RX is 
comparable to that of the brightest Hickson groups, so 
the $L_X$/$L_B$ ratio of 0.04 is  higher 
than any of the HCGs.
In contrast, the X-ray temperature lies within the 
range found for HCGs by Ponman et al (1996) and \RX lies on the
$\sigma-T$ relation for HCGs. 

The X-ray luminosity of \RX is high for its X-ray temperature, and it does not
lie on the $L_X-T$ relation of clusters or of groups. Fig. 12 shows
the $L_X-T$ relation for clusters at z$<$0.2 using data compiled from several 
sources. The solid line is the best fit using only clusters with
$T>$1 keV ($L_X\propto~T^{3.13}$; Fairley et al 1999).
Groups of galaxies with temperatures $<$1 keV show a
systematic trend to have {\it low} X-ray luminosities for a given
temperature, as already
noted by Ponman et al (1996), but \RX has the opposite property.
Infact \RX has the largest deviation above the best fit $L_X-T$ relation
of any of the clusters or groups in the plot.

\begin{figure*}
\begin{minipage}{125mm}
\psfig{figure=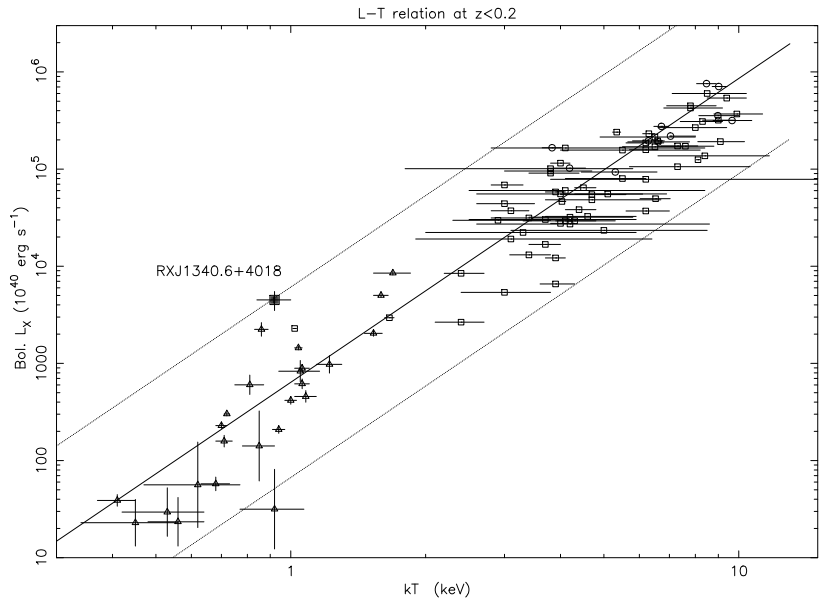,height=8.0truecm,angle=-90}
\caption{X-ray luminosity--temperature relation for clusters and groups.
\RX is the most deviant point above the best fit (solid) line. Dotted
lines are parallel to the best fit line and are merely to guide the eye.
Open squares are from David et al (1993), circles from Mushotzky \&
Scharf (1997) and triangles from Helsdon \& Ponman (1999).
}
\end{minipage}
\end{figure*}

\subsection{The luminous central galaxy}

The luminous central galaxy of \RX does not have the characteristic
envelope of cD galaxies, which would be visible as an excess
surface brightness above the r$^{1/4}$ law at $\mu\lesssim$25
mag arcsec$^{-2}$ (eg. Schombert 1986). 
Thuan \& Romanishin (1981)
found that the brightest galaxies in MKW and AWM poor clusters were
also missing cD extended envelopes. They suggested that the origin of
cD envelopes in rich clusters is tidally stripped matter 
falling on to the central galaxy, but as tidal stripping (by the cluster
potential) is
less efficient in groups, cD envelopes do not form there.

The total luminosity of galaxy G1 (M$_V$=-23.4) is similar to that 
of the brightest galaxies in the poor AWM \& MKW groups studied
by Thuan \& Romanishin (1981) and to their estimates of the 
total luminosities of cD galaxies in rich clusters, when the
cD envelope luminosities were excluded.

The lack of features in the residual images after subtraction of the
elliptical model in both the B and R bands
shows that there are no major dust lanes in the galaxy. Thus either
any dust is smoothly distributed, the progenitor galaxies were
dust-free, or any dust has been sputtered by the hot X-ray gas.

With the current X-ray data we cannot separate the X-ray luminosity
of the central galaxy from the X-ray  luminosity of the group.
Thus we cannot investigate the $L_X$/$L_B$ ratio of the galaxy alone.
The log($L_X$/$L_B$) value for the group (-1.4) is $\sim$10-100 times
larger than that for normal elliptical galaxies.
% such as NGC5846, NGC720 and NGC392.

\subsection{Environment of the group}
Because of the old ages and lack of evolution found for luminous
brightest cluster galaxies (BCGs),
it has been suggested that they may have formed before clusters, 
within groups, which then merged. It is interesting to note that
\RX lies in the outskirts of the richness class 2 cluster A1774. A1774 has a redshift of
z=0.1691 (Struble \& Rood 1987), corresponding to a line-of-sight 
velocity difference
of only 475 \kmps with \RX. 
% within the A1774 velocity dispersion of 
The projected separation is 18.2 arcmin, corresponding to 4.1 Mpc, so
\RX is in the outskirts of A1774 and is part of the 
large-scale structure associated with A1774. There is no cD galaxy
in A1774; the cluster has a Bautz-Morgan class of III, the class with least 
dominant cluster galaxies.
The centre of A1774 is south of \RX in a direction some $\approx$25\deg from the 
X-ray position angle of galaxy G1. If \RX is falling into A1774,
then the approximate alignment of the X-ray isophotes, and probably also 
the dark matter distribution, with the direction of A1774, is
suggestive of infall along a filamentary structure, as observed in 
galaxy redshift surveys and predicted in hydrodynamic 
simulations of large-scale structure 
formation (eg. Katz, Hernquist \& Weinberg 1992). 
In a detailed N-body simulation of a poor cluster within a hierarchical
cosmological model  Dubinski (1998) found that after 3 Gyr (ie. by
z$\approx$0.8) the four most massive galaxies had merged along a
primordial dark matter filament. The final luminous giant elliptical
galaxy at the cluster core had its major axis aligned along the
filament, and the appearance of the galaxies at z=0 was very similar to the
observations presented here.

\subsection{Merger age}

If the central galaxy has formed via multiple mergers, then the last major merger
probably occurred several Gyr ago, because there are now no morphological signs
left of merging activity (boxy isophotes, shells or tidal tails). The galaxy 
lies on
the fundamental plane, colour-magnitude and metallicity-
velocity dispersion relations of normal ellipticals. The time
since the last major starburst, according to the correlation of 
deviation from the fundamental plane with  spectroscopic age of Forbes
et al (1998), is $\gtrsim$4 Gyr. An estimate of the age of the stellar 
population can be made using the Lick line index system 
as defined by Worthey (1994).
Although we can not do a full
correction to the Lick system, the H$\beta$
line index is
sufficiently small (equivalent width= 0.8$\pm$0.2 \AA) to rule out a young 
age and is consistent with the
oldest ages in the Worthey models ie. 18 Gyr.

Dynamical friction causes massive galaxies to lose energy and spiral
in toward the group centre (eg. White 1976).
The timescale for dynamical friction can be estimated using the 
equations given by Binney \& Tremaine (1987) which integrate the 
Chandrasekhar (1943) dynamical friction force as a galaxy falls 
toward the group centre. Assuming the total gravitating mass 
is proportional to r$^{-2}$, consistent with our analysis of the
X-ray data, and the galaxy is initially on a circular orbit of radius
200 kpc 
with velocity $\sqrt{2}\sigma_{r}$, an L$^{*}$ galaxy with a typical 
mass/light ratio of 10 M$_{\sol}$/L$_{\sol}$ is predicted to fall 
into the group centre in $\sim$4.5x10$^{9}$ yr. Thus the lack of
observed L$^{*}$ 
galaxies within a radius of 200 kpc is consistent with the effects
of dynamical friction acting over a substantial fraction of a Hubble time.

\subsection{Comparison with simulations}

The X-ray observations of \RX require a massive dark halo to 
gravitationally bind the hot X-ray gas. The fraction of dark matter 
initially distributed in halos around individual galaxies, rather than
distributed throughout the group, is not known. Dark galaxy halos
may be stripped by galaxy interactions. 

In the simulations of virialized poor
clusters of Bode et al (1994), if 50\% of the dark matter is initially
in galaxy halos, galaxy merging occurs and a dominant central galaxy
is produced after $\sim$1x10$^{10}$ yr. Most mergers involved the dominant
galaxy. This galaxy
contains a third or more of the luminous matter in the
cluster and accretes 0.5-5$L^*$. 
% after $\sim$2x10$^{10}$ yr.
The mass function, which
was initially continuous,
develops a gap of a factor $\approx$6 in mass between the most massive
galaxy and the second most massive. 

The similarities between the properties predicted by the simulations
of Bode et al (1994) and the observed properties of \RX are striking:
the gap in the predicted mass function 
 and the gap in our observed luminosity function 
(of \m12=2.5 mag or a factor of 10), the total luminosity
($\approx$3L$^{*}$) and the high fraction 
of the total cluster light in the dominant galaxy.

In the same simulations, if all the dark matter is initially distributed
throughout the group, the initial galaxy masses are smaller, and the dynamical
friction timescales are longer.
Merging is delayed beyond 13 Gyr, 
and a central dominant galaxy is not  produced. Thus, according to
these simulations (in which 
the dark halos were 
$\sim2-4$ times more massive than the luminous matter), 
if merging is a dominant process,
the extreme properties of the \RX system could only
be produced if galaxies initially had massive dark halos.
Governato, Tozzi \& Cavaliere (1996) note that the merger rate for
galaxies with very massive dark matter
halos ($>$4 times the luminous matter mass) may not be faster
than for galaxies with less massive halos.
In the collapse phase, included in their simulations, 
unequal dark matter halos are tidally stripped 
to about the same size and mass. 

The simulations of galaxy groups of Governato, Tozzi \& Cavaliere (1996) also produced
a `merging runaway' (Cavaliere, Colafrancesco \& Menci 1991) 
in which a large fraction of the galaxies merged to
produce one or two central massive galaxies in a few crossing times
after the initial collapse of the central region. These simulations
started in a non-virialized state and included  galaxy infall.
%Compact groups were formed in $\sim$25\% of the simulations; in 
%others, with fewer infalling galaxies, 
Governato et al. (1996) also use the Press-Schechter (1974) formalism to
predict a larger fraction of systems like \RX in an open, $\Omega$=0.2
Universe than in an $\Omega$=1 Universe. In an open Universe only
10\%-15\% of groups were predicted to have formed since z=0.3, 
compared to 30\%-50\% of groups in a critical density Universe.
In the open Universe predictions there was also little galaxy infall into groups, 
and group ages corresponded to a significantly larger number
of crossing times than in a critical density Universe. 
For either value of $\Omega$, Governato et al. (1996) found that 
most galaxy merging
occurred at high redshifts z$>$0.3 (and still higher redshifts for an open Universe),
so the observed lack of morphological signs of merging is not surprising.
The detailed 
properties and statistics of systems like \RX may therefore help to 
constrain cosmological parameters
and/or  galaxy formation models.

\subsection{Formation mechanisms}

%Many studies have shown that cD galaxies in rich clusters are too 
%luminous to have been drawn from a typical Schechter LF (eg. ),
%and conclude that a special formation mechanism is required for them.
 The large gap in
 the galaxy luminosity function and the very high fraction of the
group light
emitted by the central galaxy suggest that galaxy merging has occurred,
as predicted by simulations of groups. 
At least one neighbouring galaxy (G10) has an extreme colour and may
be interacting with the central galaxy.
%The high fraction of faint neighbouring galaxies (within 200 kpc
%of the  central galaxy)
%with extreme properties 
%suggests that interactions are occurring now between
%the central galaxy and low-mass galaxies.

However, the undisturbed, relaxed morphology of 
the central galaxy, its location on the fundamental plane of
ellipticals and the lack of strong Balmer absorption lines in the
optical spectrum show that a major merger (causing major star formation
activity) has not occurred recently
(within 4 Gyr). This suggests that the majority of the merging occurred
at an early epoch, consistent with dynamical friction predictions which
would leave only low mass galaxies merging at later times.
There are other pieces of evidence that suggest the \RX system
was formed at an early epoch.  
The deviation of the system from the $L_X-T$
relation towards high X-ray luminosities can be explained in simple
terms by an early
formation epoch. At early times, the higher density of the Universe 
leads to a higher gas density and thus a higher X-ray luminosity (since
$L_X\propto$~n$^2$). 

There are puzzles remaining though. 
The lack of evidence for a  strong cooling flow is a puzzle if the
system is old, although the spatial resolution of our X-ray data may be
insufficient to resolve the central temperature decrement.
The elongation of the X-ray
isophotes suggests that the dark matter distribution is also elongated,
and thus may not be relaxed, as would be expected in an old system.
The intruiging possibility that \RX is in a filamentary structure
related to the infall of mass on to A1774 remains.

The similarity of the optical galaxy ellipticity at large radii to
the X-ray ellipticity, together with the approximate alignment of the optical
isophotes with the X-ray isophotes, supports the observed X-ray
elongation
and suggests that the stars in the outer regions of the galaxy may
be responding to the {\it group-sized} dark matter distribution.
While ellipticity increasing with radius and alignment with 
cluster structure may be a common feature of cD galaxies 
(eg. Dressler 1978), it
is rarely observed in regular giant ellipticals without cD envelopes.

\section{Conclusions}
We have measured four concordant galaxy redshifts showing that
the system \RX is a group of galaxies. ROSAT HRI X-ray observations
confirm that the X-ray emission is diffuse. The X-ray 
properties of the group are not unlike those
of the optically selected MKW and AWM poor clusters and groups,
which were selected to have dominant giant ellipticals. However,
the giant elliptical in \RX dominates the group to an extreme degree;
the second brightest galaxy is a factor of 10 less luminous (\m12=2.5
mag), and 70\% of the group optical light originates in the dominant
elliptical. This large magnitude gap, combined
with the poorness of the system, would make it difficult to detect
in two-dimensional optical surveys. 
The gap in the group galaxy luminosity function occurs $\sim$L$^*$ and 
the missing luminosity is consistent with the luminosity of the dominant galaxy.

We interpret the properties of the group in terms of merging of L$^*$
galaxies at an early epoch, leaving behind low mass galaxies, as 
predicted by dynamical friction. 
Fossil groups like 
\RX may be the long sought-after descendents of HCGs.

In this interpretation, the epoch of merging must have been early
($\gtrsim$4 Gyr ago). The undisturbed morphology of the dominant
galaxy, its location on the fundamental plane of ellipticals
and the lack of young star features in its optical spectrum all
suggest no recent merging or star formation activity. 
However, the marked elongation of the X-ray isophotes is more
consistent with a young, unrelaxed system. Future X-ray observations
will determine the detailed structure of the system as well as
any relation to the nearby cluster A1774.

Clearly more examples of groups similar to \RX need to be found 
before any firm statements about their space density can be made.
We are searching for similar systems in the WARPS X-ray selected survey 
(Scharf et al 1997, Jones et al 1998). Initial results indicate that 
there are indeed several similar systems. Detailed
observations of these fossil systems will determine the 
importance of merging in galaxy groups for
the formation of luminous elliptical galaxies.

\section{Acknowledgements}
Thanks are due to Harald Ebeling for the reduction of the Keck spectrum,
Bruce Fairley for the $L_X-T$ plot and Craig Collins for help with 
the reduction of the HRI data.
We acknowledge discussions with Robert Mann and 
use of the ING service observing programme on La Palma.
Analysis was performed on the Birmingham node of the UK Starlink
network. LRJ acknowledges support of the UK PPARC.

\section{REFERENCES}

\noindent
Albert C.E., White R.A., Morgan W.W., 1978, ApJ, 211, 309 (AWM).\\
Bahcall N.A., 1980, ApJ, 238, L117.\\
Bahcall N.A., 1981, ApJ, 247, 787.\\
Barnes J., 1985, MNRAS 215, 517.\\
Barnes J.E. 1989, Nature 338, 123.\\
Beers T.C., Flynn K., Gebhardt K., 1990, AJ, 100, 32.\\
Beers T.C., Kriessler J.R., Bird C.A., Huchra J.P. 1995, AJ, 109,
874.\\
Bender R., Burstein D., Faber S.M., 1993, ApJ, 411, 153.\\
Binney J., Tremaine S., 1987, Galactic Dynamics. Princeton Univ. Press,
 Princeton, NJ\\
Bessell M.S. 1979, PASP, 91, 543.\\
Bode P.W., Berrington R.C., Cohn H.N., Lugger P.M., 1994, ApJ,
433, 479.\\
Burstein D., Heiles C., 1984, ApJS, 54, 33.\\
Cavaliere A., Colafrancesco S., Menci N. 1991 ApJ 376, L37.\\
Chandrasekhar S., 1943, ApJ, 97, 255.\\
Christian C.A., Adams M., Barnes J.V., Butcher H., Hayes D.S.,
 Mould J.R., Siegel M., 1985, PASP, 97, 363.\\
Danese L., De Zotti G., di Tullio G., 1980, A \& A 82, 322.\\
David L.P., Slyz A., Jones C., Forman, W., Vrtilek S.D., Arnuad K.A.
1993, ApJ 412, 479.\\
Djorgovski S. 1985, PASP, 97, 1119.\\
Djorgovski S., Davis M., 1987, ApJ, 313, 59.\\
Draper P., Eaton N., 1996, "Starlink User Note 109: PISA"\\ 
Dressler A., 1978, ApJ 226, 55.\\
Dressler A., 1984, ApJ 281, 512.\\
Dubinski J., 1998, ApJ, 502, 141.\\
Faber S.M., Burstein D., Dressler A., 1977, AJ, 82, 941.\\
Fairley B.W., Jones L.R., Scharf, C.A., Ebeling, H., Perlman, E.,
 Horner, D., Wegner, G., Malkan, M. 1999, in preparation.\\
Ferguson H.C., Sandage A., 1991, AJ, 101, 765.\\
Forbes D.A., Ponman T.J., Brown R.J.N., 1998, ApJ, 508, L43.\\
Goudfrooij P., Hansen L., Jorgensen H.E., Norgaard-Nielsen H.U.,
 de Jong T., van den Hoek L.B., 1994, A\&AS 104, 179.\\
Governato F., Tozzi, P, Cavaliere A., 1996, ApJ, 458, 18.\\
Hibbard J.E., Guhathakurta P.,
 Van Gorkom J.H., Schweizer F., 1994, AJ, 107, 67.\\
Helsdon S., Ponman, T.J., 1999, MNRAS submitted.\\
Hunsberger S.D., Charlton J.C., Zaritsky D., 1998, ApJ, 505, 536.\\
Jedrzejewski R.I., 1987, MNRAS, 226, 747.\\
Jenkins E.B., Savage B.D., 1974, ApJ, 187, 243.\\
Jones L.R., Fong R., Shanks T., Ellis R.S., Peterson B.A. 1991,
MNRAS, 249, 481.\\
Jones L.R., Scharf C.A., Ebeling H., Perlman E., Wegner G.,
 Malkan M., Horner, D., 1998, ApJ, 495, 100.\\
Jones L.R. et al. 1999, in preparation.\\
Katz N., Hernquist L., Weinberg D.H., 1992, ApJ, 399, L109.\\
Kriss G.A., Cioffi D.F., Canizares C.R., 1983, ApJ, 272, 439.\\
Kron R.G. 1980 ApJS, 43, 305.\\
Landolt A.U., 1992. AJ, 104, 340.\\
Metcalfe N., Shanks T., Fong R., Jones L.R., 1991, MNRAS, 249, 
   498.\\
Metcalfe N., Shanks T., Fong R., Roche N., 1995, MNRAS, 273, 257.\\
Morgan W.W., Kayser S., White R.A., 1975, ApJ, 199, 545 (MKW).\\
Mulchaey J.S., Davis D.S., Mushotzky R.F., Bursstein D., 1996,
ApJ, 456, 80.\\
Mulchaey J.S., Zabludoff, A.I., 1999, ApJ, 514, 133.\\
Mushotzky R.F., Scharf, C.A. 1997 ApJ, 482, L13.\\
Oemler A. 1976, ApJ 209, 693.\\
Pence W. 1976, ApJ, 203, 39.\\
Prandoni I., Iovino A., MacGillivray H.T., 1994, AJ 107, 1235.\\
Price R., Burns J.O., Duric N., Newberry M.V., 1991, AJ, 102, 14.\\
Ponman T.J., Bertram D., 1993, Nature, 363, 51.\\
Ponman T.J., Allan D.J., Jones L.R., Merrifield M., McHardy I.M.,
 Lehto H.J., Luppino G.A. 1994, Nature, 369, 462 (P94). \\
Ponman T.J., Bourner, P.D.J., Ebeling, H., Bohringer H., 1996, MNRAS,
 283, 690.\\
Press W.H., Schechter P., 1974, ApJ, 187, 425.\\
Prugniel Ph., Simien F., 1996, A\&A, 309, 749.\\
Ramella M., Geller M.J., Huchra J.P. 1989, ApJ 344, 57.\\
Saglia R.P. et al. 1997, MNRAS, 292, 499.\\
Sandage A., Binggeli B., Tammann G.A., 1985, AJ, 90, 1759.\\
Scharf C.A., Jones L.R., Ebeling H., Perlman E., Malkan M.,
 Wegner G. 1997, ApJ, 477, 79.
Schombert J.M. 1986, ApJS, 60, 303.\\
Smail I., Hogg D.W., Yan L., Cohen J.G., 1995, ApJ, 449, L105.\\
Steidel C.C., Hamilton D., 1993, AJ, 105, 2017.\\
Struble M.F. Rood H.J. 1987, ApJS, 63, 543.\\
Thuan T.X., Romanishin W., 1981, ApJ 248, 439.\\
Vikhlinin A., McNamara B.R., Hornstrup A., Quintana H., Forman W.,
Jones C., Way C., 1999, ApJ submitted.\\
White D.A., Jones C., Forman W., 1997, MNRAS 292, 419.\\
White S.D.M., 1976, MNRAS, 174, 19.\\
Worthey G., 1994, ApJS, 95, 107.\\
Yamagata T., Maehara H. 1986, PASJ, 38, 661.\\
Zabludoff A.I, Mulchaey J.S., 1998, ApJ 496, 39.\\

\end{document}